\documentclass[12pt]{article}

\usepackage[letterpaper,left=1.25in,right=1.25in,top=1in,bottom=1in]{geometry}

\usepackage[T1]{fontenc}
\usepackage[utf8]{inputenc}
\usepackage{newtxtext,newtxmath} 
\usepackage{graphicx}

\usepackage{setspace}
\usepackage{titlesec}
\titleformat{\section}{\bfseries\fontsize{16}{18}\selectfont}{\thesection}{1em}{}
\titlespacing*{\section}{0pt}{2\baselineskip}{1\baselineskip}

\titleformat{\subsection}{\bfseries\fontsize{14}{16}\selectfont}{\thesubsection}{1em}{}
\titlespacing*{\subsection}{0pt}{2\baselineskip}{1\baselineskip}

\titleformat{\subsubsection}{\bfseries\fontsize{12}{14}\selectfont}{\thesubsubsection}{1em}{}
\titlespacing*{\subsubsection}{0pt}{2\baselineskip}{1\baselineskip}

\usepackage{fancyhdr}
\usepackage{graphicx}
\usepackage{float}
\usepackage{booktabs}
\usepackage{caption}
\usepackage[numbers]{natbib}

\usepackage[colorlinks=true,linkcolor=black,citecolor=black,urlcolor=black]{hyperref}

\newcommand{\MICSTitle}[1]{%
  \begin{center}
    \vspace*{1.5in} 
    {\fontsize{18}{20}\selectfont #1\par}
    \vspace{2\baselineskip} 
  \end{center}
}

\newcommand{\MICSAuthorBlock}[5]{%
  \begin{center}
    {\fontsize{14}{16}\selectfont #1\par} 
    {\fontsize{14}{16}\selectfont #2\par} 
    {\fontsize{14}{16}\selectfont #3\par} 
    {\fontsize{14}{16}\selectfont #4\par} 
    {\fontsize{14}{16}\selectfont #5\par} 
    \vspace{2\baselineskip} 
  \end{center}
}

\newcommand{\MICSAbstract}[1]{%
  \begin{center}
  \newpage
    {\bfseries\fontsize{16}{18}\selectfont Abstract\par}
  \end{center}
  \vspace{\baselineskip} 
  {\fontsize{12}{14}\selectfont
  #1\par}
}

\newcommand{\MICSIntro}[1]{%
  \begin{center}
    {\bfseries\fontsize{16}{18}\selectfont Intro\par}
  \end{center}
  \vspace{\baselineskip} 
  {\fontsize{12}{14}\selectfont
  #1\par}
}

\newcommand{\MICSRelatedWork}[1]{%
  \begin{center}
    {\bfseries\fontsize{16}{18}\selectfont Related Work\par}
  \end{center}
  \vspace{\baselineskip} 
  {\fontsize{12}{14}\selectfont
  #1\par}
}

\newcommand{\MICSImplementation}[1]{%
  \begin{center}
    {\bfseries\fontsize{16}{18}\selectfont Implementation\par}
  \end{center}
  \vspace{\baselineskip} 
  {\fontsize{12}{14}\selectfont
  #1\par}
}

\newcommand{\MICSExperimentation}[1]{%
  \begin{center}
    {\bfseries\fontsize{16}{18}\selectfont Experimentation\par}
  \end{center}
  \vspace{\baselineskip} 
  {\fontsize{12}{14}\selectfont
  #1\par}
}

\newcommand{\MICSResultsAnalysis}[1]{%
  \begin{center}
    {\bfseries\fontsize{16}{18}\selectfont Results and Analysis\par}
  \end{center}
  \vspace{\baselineskip} 
  {\fontsize{12}{14}\selectfont
  #1\par}
}

\newcommand{\MICSConclusion}[1]{%
  \begin{center}
    {\bfseries\fontsize{16}{18}\selectfont Conclusion\par}
  \end{center}
  \vspace{\baselineskip} 
  {\fontsize{12}{14}\selectfont
  #1\par}
}
\begin{document}
\thispagestyle{empty}

\MICSTitle{The Effect of Text Chunk Size on Retrieval-Augmented Generation Performance}

\MICSAuthorBlock
  {German Garrido-Lestache Belinchon , Hugo Garrido-Lestache Belinchon}
  {Department of Computer Science and Software Engineering}
  {Milwaukee School of Engineering}
  {Milwaukee, WI US}
  {garrido-lestachebeli@msoe.edu}

\MICSAbstract{%
Retrieval-Augmented Generation (RAG) systems have emerged as a powerful process for allowing large language models (LLMs) to retrieve relevant information to use as source material during text generation. A critical yet under-explored component of these systems is the granularity at which source documents are segmented into retrievable chunks. The size of these chunks has the potential to significantly influence generation quality, contextual correctness, retrieval precision, and computational efficiency. Despite its importance, chunk size is often selected without proper evaluation of its impact on generation quality.

This paper aims to explore the effect that text size has on generation quality by using multiple representations of identical full-length undergraduate textbooks, each split into different sized chunks, such as sentences, paragraphs, or chapters. As the same source material is used for all the chunks, it isolates chunk granularity as the primary variable.  These chunks are embedded using a dense embedding model and indexed in a vector database for semantic retrieval. Relevance is determined through cosine similarity search, where the most similar text chunks to a user’s prompt are retrieved and supplied as contextual input to the LLM in order to produce a response.

Smaller chunks, such as individual sentences, may allow for precise retrieval by narrowing the focus of each chunk. However, they contain less information, which may limit the model’s ability to generate coherent responses. Larger chunks, such as entire chapters, contain lots of broad information that may improve correctness, but also introduce additional noise and increase computational cost. Because larger chunks contain more information, the number of chunks returned to the model must also be considered.

This paper evaluates how chunk size, along with the number of retrieved segments, influences generation quality and retrieval effectiveness. By comparing these configurations, this study seeks to better understand how document segmentation affects the performance and efficiency of Retrieval-Augmented Generation systems.
}

\MICSIntro{%
LLMs are incredibly effective at generating coherent, relevant text in a number of different formats and contexts. Though the text they generate may seem to fit at first glance, upon closer inspection its often riddled with outdated, made up, and often incorrect information. This is a well known issue of LLMs, but it would be incredibly time consuming and computationally expensive to train them on every conceivable topic, so how do we make these models produce factual, intelligent responses.

Retrieval Augmented Generation (RAG) systems address this limitation by combining external knowledge retrieval with text generation. These models will use a large database to store chunks of informational text, and a  RAG model will then, given a prompt, find relevant source information and supply them during text generation to produce accurate factual responses. This pipeline of course relies heavily on the model's ability to retrieve relevant information.

Before documents are embedded and indexed into a vector database, they must be segmented into retrievable chunks. The granularity of these chunks presents a tradeoff. Smaller chunks increase retrieval speed, and computational power, but lack the contextual depth for coherent responses. Larger chunks may contain more context for the LLM to reference, but introduce noise, decrease retrieval efficiency, and increases computational overhead. The size of these chunks, be they sentences or paragraphs, is often selected without proper evaluation of how this decision impacts overall generation quality.

Chunk size has a clear impact on context retrieval, and coherent text generation. Despite this, few studies isolate chunk size as a variable. This paper aims to address that by evaluating how text chunk granularity affects retrieval effectiveness and generation quality in a RAG framework. By comparing multiple chunk sizes with a single identical source material, this work aims to provide insight on an important, yet often overlooked component of Retrieval Augmented Generation systems.
}

\MICSRelatedWork{%
Retrieval-Augmented Generation has become a foundational approach for improving response accuracy in large language models by using external sources during generation. Early RAG solutions focus on retrieval architectures and embedding quality, while keeping a fixed document segmentation strategy (Lewis, et al., 2020) \cite{rag}. Retrieval-focused generation models such as Fusion-in-Decoder further demonstrate that integration between retrieval and generation can significantly enhance response quality, yet continue to treat document chunking as a static preprocessing step (Izacard et al., 2021) \cite{retrievalfocused}.

More recent work has improved how retrieved information is structured and utilized. The RAPTOR framework (Sarthi et al., 2024) \cite{abstractiveprocessing} introduces a hierarchical retrieval, based on recursive summarization over large document collections. Similarly, Self-RAG (Asai et al., 2023) \cite{retrievegenerate} enables an iterative generation-critique loop, allowing models to refine both retrieval decisions and generated outputs. These approaches highlight that improvements to retrieval and generation pipelines can substantially enhance RAG system performance.

Other research addresses retrieval noise and context relevance through post-retrieval refinement. Corrective RAG (Yan et al., 2024) \cite{correctiverag} introduces filtering and correction mechanisms to mitigate the impact of irrelevant retrieved content, while ChunkRAG (Singh et al., 2024) \cite{chunkrag} applies LLM-based relevance filtering prior to generation. Although effective, these approaches largely operate after retrieval and assume a fixed segmentation of sources, focusing on refining retrieved content rather than examining how document structure influences performance.

More closely related to this work are studies that examine document chunking within RAG pipelines. (Hladena et al., 2025) \cite{chunksize} demonstrate that chunk size significantly impacts retrieval efficiency, generation quality, and computational cost. Additional studies analyse chunk size across datasets and embeddings (Bhat et al., 2025) \cite{rethinkingchunks} or leverage document structure in domain-specific contexts (Yepes et al., 2025) \cite{financialchunking}, though these works primarily emphasize retrieval effectiveness. In contrast, my work isolates chunk size as the primary variable using meaningful boundaries (sentences, paragraphs, pages, and chapters) from identical source content and evaluated under controlled embedding, retrieval, and generation conditions. This allows us to examine both retrieval effectiveness and downstream generation quality, providing a view of how document chunking influences RAG system performance.
}

\MICSImplementation{%
The RAG pipeline begins by loading and extracting the text, page by page, from the source document. It extracts the raw text and removes artifacts such as, excessive line breaks, page numbers, and hyphenated word splits. This produces a clean continuous body of text suitable for chunking.

The document is then segmented using different levels of granularity, including chapter-level splitting using regex, as well as agentic paragraph and sentence level segmentation, in which an LLM iteratively extracts the first segment from a sliding window of text. This allows for more natural segmentation rather than relying purely on formatting which allowed the chunking to work on any medium of source document.

Once segmented, each chunk is embedded into a vector representation, using a pre-trained embedding model, and stored into a separate vector database for each chunk granularity. During retrieval, a user's prompt is embedded with the same model, and a cosine similarity search is used to find the n most similar text chunks.

The system then concatenates the retrieved chunks, to be used as source information, along with the user's query. This augmented prompt is then passed onto an LLM to generate a factual, grounded response.

Overall, this implementation enables controlled experimentation across different chunk sizes and retrieval configurations, allowing for direct comparison of their impact on generation quality, relevance, and efficiency. 

\begin{figure}[H]
  \centering
  \includegraphics[width=1\linewidth]{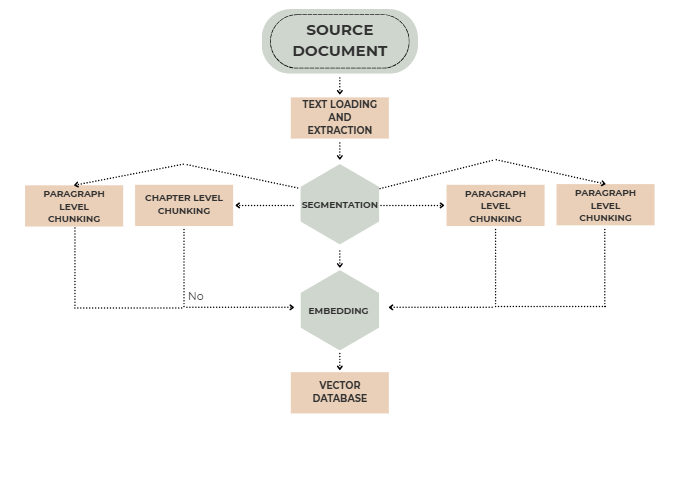}
  \caption{Flowchart Depicting the Source Document Pipeline}
  \label{fig:example}
\end{figure}
\begin{figure}[H]
  \centering
  \includegraphics[width=0.25\linewidth]{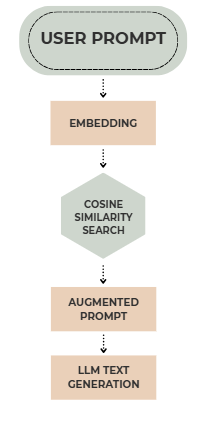}
  \caption{Flowchart Depicting the RAG Pipeline}
  \label{fig:example}
\end{figure}
}

\MICSExperimentation{%
To first ensure the program could effectively retrieve relevant source chunks, I produced a set of 100 prompts, generated by ChatGPT, that were intended to simulate realistic prompts, varying in scope specificity and complexity. For each prompt, I tested the retrieval across all chunking strategies. The retrieval effectiveness was evaluated using Reciprocal Rank, which measures how high the first relevant result appears within a ranked list of retrieved chunks. A retrieved chunk was considered relevant if it contained the information necessary to correctly answer the prompt. Relevance was determined agentically, using an LLM to  verifying whether the answer could be found within the retrieved text. This ensured consistency across all chunking strategies. By averaging the reciprocal rank scores across all the prompts, I obtained a retrieval score that I could compare for each chunking method. Additionally, I graphed the score of each chunking strategy for each prompt to analyse which specific prompts the model struggles to retrieve.
}

\MICSResultsAnalysis{%
\begin{figure}[H]
  \centering
  \includegraphics[width=0.75\linewidth]{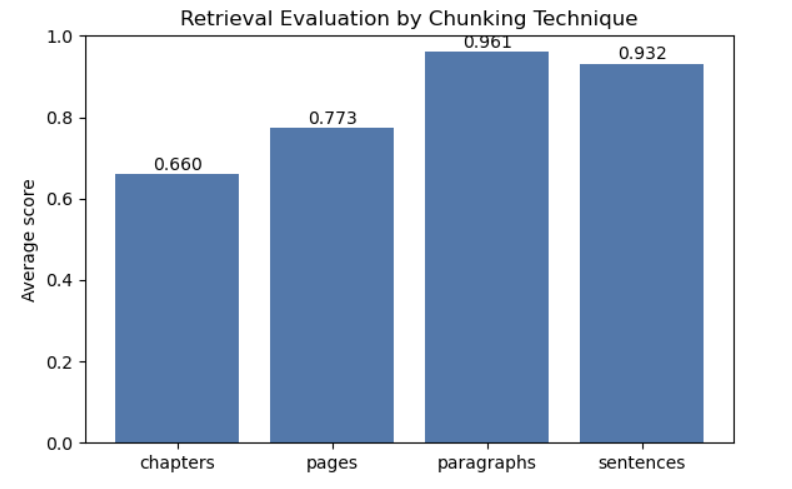}
  \caption{Results from Chunking a Mathematics Textbook}
  \label{fig:example}
\end{figure}
\begin{figure}[H]
  \centering
  \includegraphics[width=0.75\linewidth]{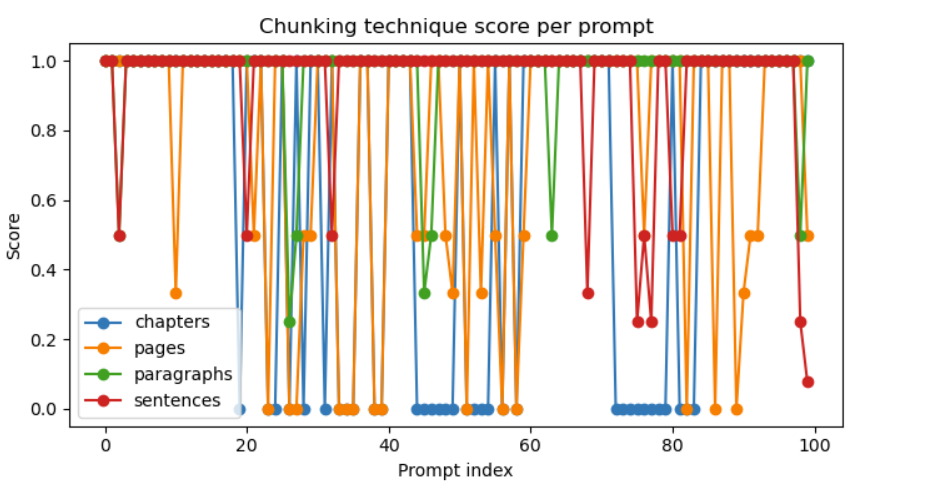}
  \caption{Results from Chunking a Mathematics Textbook}
  \label{fig:example}
\end{figure}
The results demonstrate that chunking strategy has a significant impact on retrieval performance, though this varies depending on the structure of the underlying text. For a mathematics textbook, paragraph-level chunking achieved the highest average retrieval score, followed closely by sentence-level chunking. This suggests that for structured, information-dense content, medium-sized chunks such as paragraphs provide the best balance between precision and contextual completeness. Sentence-level chunks, while precise, occasionally lacked sufficient context, whereas chapter-level chunks were too broad, introducing noise and reducing retrieval accuracy.

\begin{figure}[H]
  \centering
  \includegraphics[width=0.75\linewidth]{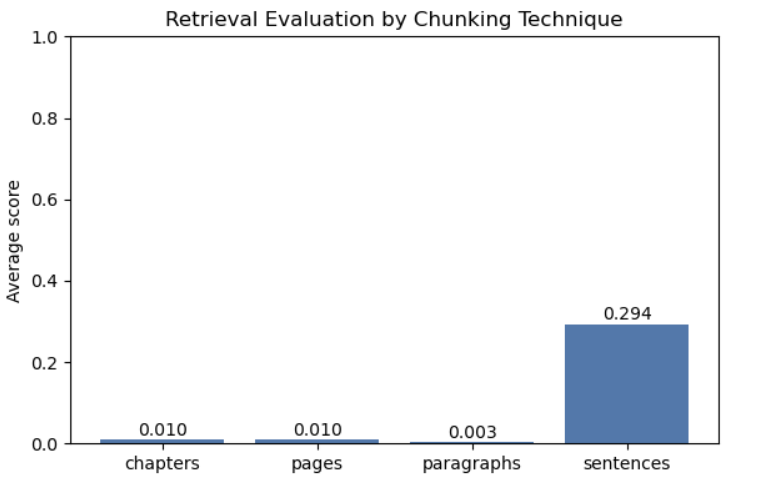}
  \caption{Results from Chunking a Narrative Text}
  \label{fig:example}
\end{figure}
\begin{figure}[H]
  \centering
  \includegraphics[width=0.75\linewidth]{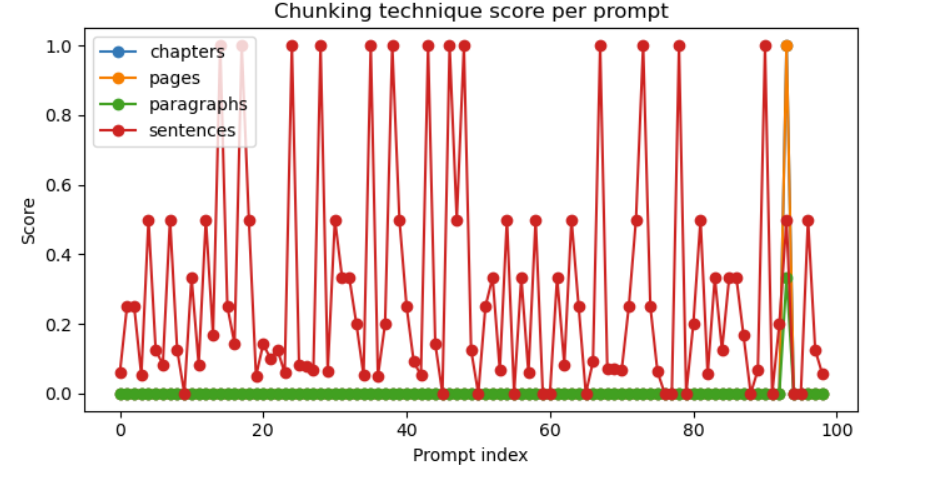}
  \caption{Results from Chunking a Narrative Text}
  \label{fig:example}
\end{figure}
The results for the narrative text show a very different pattern. Sentence-level chunking significantly outperformed all other methods, achieving an average retrieval score of 0.294, while chapter, page, and paragraph-level chunking all performed extremely poorly. This indicates that for narrative-driven content, where relevant details are often localized to specific lines of dialogue or description, smaller chunks are far more effective. Larger chunks in this context appear to dilute relevant information, making it more difficult for the retrieval system to isolate useful content. These results reinforce the conclusion that optimal chunk size is highly dependent on the structure and writing style of the source material.

Despite these findings, several limitations must be considered. While smaller chunk sizes often improve retrieval precision, they result in a significantly larger number of stored embeddings, increasing both storage requirements and indexing time. On the other hand, larger chunks reduce storage overhead but may introduce irrelevant context, negatively impacting retrieval quality and introducing unwanted noise during generation. Furthermore, the use of agentic chunking introduces additional computational cost, as it requires a large language model during preprocessing. This makes it less practical for large-scale systems despite its potential to produce more meaningful text chunks.
}

\MICSConclusion{%
This study set out to investigate how document segmentation chunk size influences the performance of RAG systems. The results demonstrate that chunky strategy plays a meaningful role in retrieval and generation. However, no single chunk size consistently outperforms others across all mediums. Instead, the effectiveness of a given chunk size is tied to the structure of the text.

One key finding is that different text mediums exhibit different optimal chunking behaviours. More structured or narrative-driven documents tend to benefit from larger chunks, where broader context improves the model’s ability to generate coherent and accurate responses. In contrast, more fragmented or information-dense texts may benefit from smaller chunks, which allow for more precise retrieval at the cost of reduced contextual completeness. This highlights an important trade-off between precision and context, suggesting that chunk size should not be treated as a fixed parameter, but rather as a configurable aspect of system design.

These results point to several promising avenues for future research. One particularly compelling direction is the further exploration of agentic or adaptive chunking, where the chunks are dynamically segmented based on document structure, query intent, or retrieval feedback. Such systems could adjust chunk granularity at indexing time or retrieval time, selecting smaller chunks for fact-based queries and larger, more contextual segments for reasoning-heavy prompts. Additionally, hybrid approaches that combine multiple chunk sizes or assemble context dynamically from atomic units may further mitigate the precision–context trade-off observed in static chunking schemes.

Overall, this work reinforces the idea that effective RAG systems require careful consideration of how source documents are segmented, rather than relying on a one-size-fits-all approach. By better aligning chunking methods with the structure of the source material, it is possible to improve both retrieval quality and generation performance in a more consistent and efficient manner.
}

\newpage
\pagestyle{fancy}
\bibliographystyle{plainnat}
\bibliography{references}

@misc{abstractiveprocessing,
      title={RAPTOR: Recursive Abstractive Processing for Tree-Organized Retrieval}, 
      author={Parth Sarthi and Salman Abdullah and Aditi Tuli and Shubh Khanna and Anna Goldie and Christopher D. Manning},
      year={2024},
      eprint={2401.18059},
      archivePrefix={arXiv},
      primaryClass={cs.CL},
      url={https://arxiv.org/abs/2401.18059}, 
}

@misc{retrievegenerate,
      title={Self-RAG: Learning to Retrieve, Generate, and Critique through Self-Reflection}, 
      author={Akari Asai and Zeqiu Wu and Yizhong Wang and Avirup Sil and Hannaneh Hajishirzi},
      year={2023},
      eprint={2310.11511},
      archivePrefix={arXiv},
      primaryClass={cs.CL},
      url={https://arxiv.org/abs/2310.11511}, 
}

@misc{correctiverag,
      title={Corrective Retrieval Augmented Generation}, 
      author={Shi-Qi Yan and Jia-Chen Gu and Yun Zhu and Zhen-Hua Ling},
      year={2024},
      eprint={2401.15884},
      archivePrefix={arXiv},
      primaryClass={cs.CL},
      url={https://arxiv.org/abs/2401.15884}, 
}

@inproceeding{chunksize,
  title     = {The Effect of Chunk Size on the RAG Performance},
  author    = {Hladěna, Jan and Šteflovič, Kirsten and Čech, Pavel and Štekerová, Kamila and Žváčková, Andrea},
  booktitle = {Software Engineering: Emerging Trends and Practices in System Development},
  year      = {2025},
  publisher = {Springer}
}

@article{rag,
  title   = {Retrieval-Augmented Generation for Knowledge-Intensive NLP Tasks},
  author  = {Lewis, Patrick and Perez, Ethan and Piktus, Aleksandra and others},
  journal = {Advances in Neural Information Processing Systems},
  year    = {2020}
}

@inproceedings{retrievalfocused,
  title     = {Leveraging Passage Retrieval with Generative Models for Open Domain Question Answering},
  author    = {Izacard, Gautier and Grave, Edouard},
  booktitle = {ACL},
  year      = {2021}
}

@article{financialchunking,
  title   = {Financial Report Chunking for Effective Retrieval Augmented Generation},
  author  = {Jimeno Yepes, Antonio and others},
  journal = {arXiv preprint arXiv:2402.05131},
  year    = {2024}
}

@article{rethinkingchunks,
  title   = {Rethinking Chunk Size for Long-Document Retrieval: A Multi-Dataset Analysis},
  author  = {Bhat, Sinchana Ramakanth and others},
  journal = {arXiv preprint arXiv:2505.21700},
  year    = {2025}
}

@article{chunkrag,
  title   = {ChunkRAG: Novel LLM-Chunk Filtering Method for RAG Systems},
  author  = {Singh, Ishneet Sukhvinder and others},
  journal = {arXiv preprint arXiv:2410.19572},
  year    = {2024}
}
\end{document}